\documentclass[apj]{emulateapj}

\tightenlines

\usepackage{ulem} 
\usepackage{color} 
\usepackage{graphicx}
\usepackage{epsfig,amsmath}
\usepackage{times}

\makeatletter

\newcommand{\Rmnum}[1]{\expandafter\@slowromancap\romannumeral #1@}
\makeatother

\newcount\listnorom
\newcommand\listromanDE{\global\advance \listnorom by 1
{\lowercase\expandafter{(\romannumeral\listnorom)}\ }}
\newcommand\newlistroman{\listnorom=0}

\def\lsim{\raise0.3ex
  \hbox{$<$\kern-0.75em\raise-1.1ex\hbox{$\sim$}}\,}
\def\gsim{\raise0.3ex
  \hbox{$>$\kern-0.75em\raise-1.1ex\hbox{$\sim$}}\,}

\newcommand{\fHe}{f_\mathrm{He}}
\newcommand{\fSN}{f_\mathrm{SN}}
\newcommand{\MB}{Maxwell-Boltzmann}
\newcommand{\Pgas}{P_\mathrm{gas}}
\newcommand{\Msonic}{M_{S,0}}
\newcommand{\Malf}{M_{A,0}}

\newcommand{\fDamp}{f_\mathrm{damp}}
\newcommand{\Nphot}{N_{\mathrm{phot}}}
\newcommand{\fzCR}{f_1^{\mathrm{cr}}(p)}
\newcommand{\crPup}{\xi_\mathrm{CR}}
\newcommand{\crPgas}{\chi_\mathrm{CR}}

\newcommand{\crhydro}{{\it cr-hydro-NEI}}

\newcommand{\Fermi}{{\it Fermi-LAT}}
\newcommand{\Suz}{{\it Suzaku}}
\newcommand{\HESS}{{\it HESS}}

\newcommand{\Inj}{\chi_\mathrm{inj}}

\newcommand{\SunMyr}{$\Msun$\,yr$^{-1}$}

\newcommand{\FS}{forward shock}
\newcommand{\RS}{reverse shock}

\newcommand{\pmax}{p_\mathrm{max}}

\newcommand{\Norm}{f_\mathrm{norm}} 

\newcommand{\dMdt}{dM/dt}
\newcommand{\Vwind}{V_\mathrm{wind}}
\newcommand{\Twind}{T_\mathrm{wind}}
\newcommand{\SigWind}{\sigma_\mathrm{wind}}
\newcommand{\cutoff}{\alpha_\mathrm{cut}}
\newcommand{\Bcut}{\beta_\mathrm{cut}}

\newcommand{\gtsim}{\mathrel{\hbox{\raise0.2ex \hbox{$>$}\kern-0.75em\raise-0.9ex\hbox{$\sim$}}}}

\newcommand{\Ptot}{P_\mathrm{tot}}

\newcommand{\pcc}{cm$^{-3}$}
\newcommand\Msun{\mathrm{M}_{\odot}}

\newcommand{\Bamp}{B_\mathrm{amp}}
\newcommand{\EffDSA}{{\cal E_\mathrm{DSA}}}

\newcommand{\EffEsc}{\cal E_\mathrm{esc}}
\newcommand{\Kep}{K_\mathrm{ep}}

\newcommand\tSNR{t_\mathrm{SNR}}
\newcommand\dSNR{d_\mathrm{SNR}}
\newcommand\EnSN{E_\mathrm{SN}}
\newcommand\Mej{M_\mathrm{ej}}
\newcommand{\RFS}{R_\mathrm{FS}}

\newcommand{\VelFS}{V_\mathrm{FS}}

\newcommand{\SA}{semi-analytic}

\newcommand{\NEI}{non-equilibrium ionization}

\newcommand{\xx}[1]{\!\times\!10^{#1}}
\newcommand{\Tacc}{t_\mathrm{acc}}

\newcommand{\DSA}{diffusive shock acceleration}
\newcommand{\CSM}{circumstellar medium}
\newcommand{\MFA}{magnetic field amplification}

\newcommand{\kmps}{km s$^{-1}$}
\newcommand{\NL}{nonlinear}
\newcommand{\gamray}{$\gamma$-ray}

\newcommand{\fFEB}{f_\mathrm{feb}}
\newcommand{\Valf}{v_A}
\newcommand{\alf}{Alfv\'en}
\newcommand{\Alf}{Alfv\'en}
\newcommand{\fAlf}{f_\mathrm{alf}}
\newcommand{\falf}{f_\mathrm{alf}}
\newcommand{\SNRJ}{SNR RX J1713.7-3946}

\newcommand{\SNRJmm}{SNR J1713}
\newcommand{\SC}{self-consistent}
\newcommand{\SCly}{self-consistently}
\newcommand{\Rtot}{R_\mathrm{tot}}
\newcommand{\Rsub}{R_\mathrm{sub}}
\newcommand{\muG}{$\mu$G}

\newcommand{\be}{\begin{eqnarray}}
\newcommand{\ee}{\end{eqnarray}}

\newcommand{\rel}{relativistic}
\newcommand{\nonrel}{non\-rel\-a\-tiv\-is\-tic}

\newcommand{\TP}{test-particle}

\newcommand{\mc}{Monte Carlo}
\newcommand{\MC}{Monte Carlo}
\newcommand{\syn}{synchrotron}

\newcommand{\pion}{pion-decay}
\newcommand{\IC}{inverse-Compton}
\newcommand{\brems}{bremsstrahlung} 

\newcommand{\mrm}{\mathrm}
\newcommand{\tna}{\,\tablenotemark{a}}
\newcommand{\tnb}{\,\tablenotemark{b}}
\newcommand{\tnc}{\,\tablenotemark{c}}
\newcommand{\tnd}{\,\tablenotemark{d}}
\newcommand{\tne}{\,\tablenotemark{e}}
\newcommand{\tnf}{\,\tablenotemark{f}}
\newcommand{\tng}{\,\tablenotemark{g}}

\newcommand{\tnaa}{\,\tablenotemark{$\dagger$}}
\newcommand{\tnab}{\,\tablenotemark{$\ddagger$}}
\newcommand{\tnac}{\,\tablenotemark{$\ast$}}

\newcount\Inum
\newcount\IInum
\newcount\IIInum
\newcount\IVnum
\def\I{\global\multiply\IInum by 0 \global\multiply\IIInum by 0
            \global\multiply\IVnum by 0 \global\advance \Inum by 1
            {\the\Inum. }}
\def\II{\global\multiply\IIInum by 0\global\multiply\IVnum by 0
       \global\advance \IInum by 1 {\the\Inum.\the\IInum. }}
\def\III{\global\multiply\IVnum by 0\global\advance \IIInum by 1
            {\the\Inum.\the\IInum.\the\IIInum. }}
\def\IV{\global\advance \IVnum by 1
            {\the\IVnum. }}
\begin{document}

\title{A Generalized Model of Nonlinear Diffusive Shock Acceleration Coupled to an Evolving Supernova Remnant}

\author{Shiu-Hang Lee,\altaffilmark{1}
Donald C. Ellison\altaffilmark{2}
 and Shigehiro Nagataki\altaffilmark{1}}

\altaffiltext{1}{Yukawa Institute for Theoretical Physics, Kyoto University Oiwake-cho Kitashirakawa, Sakyo-ku, Kyoto 606-8502, Japan;
lee@yukawa.kyoto-u.ac.jp; nagataki@yukawa.kyoto-u.ac.jp}

\altaffiltext{2}{Physics Department, North Carolina State
University, Box 8202, Raleigh, NC 27695, U.S.A.;
don\_ellison@ncsu.edu}

\begin{abstract}
To better model the efficient production of cosmic rays (CRs) in supernova remnants (SNRs) with the associated coupling between CR production and SNR dynamics, we have generalized an existing \crhydro\ code \citep[i.e.,][]{ESPB2012} to include the following processes:
(1) an explicit calculation of the upstream precursor structure including the position dependent flow speed, density, temperature, and magnetic field strength; (2) a momentum and space dependent CR diffusion coefficient; (3) an explicit calculation of magnetic field amplification (MFA); (4) calculation of the maximum CR momentum using the amplified magnetic field; (5) a finite \Alf\ speed for the particle scattering centers; and (6) the ability to accelerate a superthermal seed population of CRs as well as the ambient thermal plasma.
While a great deal of work has been done modeling SNRs, most work has concentrated on either the continuum emission from relativistic electrons or ions, or the thermal emission from the shock heated plasma. 
Our generalized code combines these elements and 
describes the interplay between CR production and SNR evolution, including the \NL\ coupling of efficient diffusive shock acceleration (DSA), based mainly on the work of 
P. Blasi and co-workers, and a non-equilibrium ionization (NEI) calculation of thermal X-ray line emission.
We believe our generalized model will provide a consistent modeling platform for SNRs, including those interacting with molecular clouds, and improve the interpretation of current and future observations, including the high-quality spectra expected from   {\it Astro-H}.
\SNRJ\ is modeled as an example.
%
\end{abstract}

\keywords{acceleration of particles, shock waves, ISM: cosmic rays,
           ISM: supernova remnants, magnetic fields, turbulence}

\section{Introduction}
In this paper we report on an extensive generalization of a 
\crhydro\ code \citep[e.g.,][]{EPSBG2007,PES2009,PSRE2010, EPSR2010,ESPB2012} that has been used to model cosmic-ray (CR) production in evolving supernova remnants (SNRs). This code 
uses a \SA, \NL\ \DSA\ (DSA) model to calculate the CR production at the remnant \FS\ (FS) and determines the broadband continuum radiation from CR electrons and ions consistently with the thermal X-ray line emission from the shock-heated plasma. The feedback effects from efficient CR production influence the SNR dynamics and the \NEI\ (NEI) calculation of the thermal emission.
Until now, this code has used the \NL\ DSA calculation described in \citet*{BGV2005}. 
In a series of papers since then 
\citep[e.g.,][and references therein]{CBAV2009}, Blasi and 
co-workers have presented a number of important generalizations to their \SA\ model and we have included many of these, as well as additional effects, in an updated version of the \crhydro\ code.

We note that in addition to the work of Blasi and 
co-workers, a great deal of work on \NL\ DSA has been done by a large number of researchers in the last 30 years
\citep[for reviews see][]{Drury83,BO78,JE91,MD2001,SchureEtal2012}. A good description of some of the numerical techniques used is given in \citet{CKVJ2010} where a distinction is made between fully numerical time-dependent solutions \citep[e.g.,][]{Bell1987,BV97,KJ2009,ZirA2010}, stationary \mc\ simulations \citep*[e.g.,][]{JE91,VBE2009}, and quasi-stationary \SA\ solutions \citep*[e.g.,][]{MDV2000,BAC2007,CAB2010}.
In addition to these techniques there are direct 
particle-in-cell (PIC) plasma simulations that, in principle, can model all aspects of shock formation, particle injection and acceleration, and the 
self-generation of magnetic turbulence 
\citep[e.g.,][]{HoshinoSurf2002,RiqSpit2010,RiqSpit2011,GarSpit2012}
\citep*[see, however,][for a discussion of the difficulties in applying PIC simulations  to SNRs]{VBE2008}. 
We have chosen to adapt the \SA\ Blasi model to our \crhydro\ code because this model contains most of the essential \NL\ effects we need, is computationally fast, and is fully documented in published papers.


\newlistroman
The specific generalizations we include are:
\noindent\listromanDE\ the explicit calculation of the upstream precursor structure including the position dependent flow speed, density, temperature, and magnetic field strength;
\noindent\listromanDE\ a momentum and space dependent CR diffusion coefficient;
\noindent\listromanDE\ an explicit calculation of \MFA\ (MFA) that replaces our previous ad hoc parameterization;
\noindent\listromanDE\ the calculation of the maximum CR 
momentum, $\pmax$, using the amplified magnetic field; 
\noindent\listromanDE\ a finite \alf\ speed for the particle scattering centers;
\noindent\listromanDE\ the ability to accelerate a superthermal seed population of CRs as well as the ambient thermal plasma;
\noindent\listromanDE\ a calculation of the break-momentum and 
high-energy cutoff of the CR distribution due to ion-neutral damping when the FS interacts with a partially ionized medium; and 
\noindent\listromanDE\ Coulomb losses for low energy but still superthermal CRs.

Many assumptions and approximations are required for these generalizations and we detail them here. 
We compare results from our new generalized code to previous work modeling \SNRJ\ 
(henceforth \SNRJmm) and point out where significant changes have occurred. While our generalized model shows  differences from the description of \SNRJmm\  given in \citet{ESPB2012}, our most important conclusion, that the GeV-TeV emission is dominated by \IC\ 
(IC) emission from \rel\ electrons rather than \pion\ emission from \rel\ ions, remains robust. 
The fit to \SNRJmm\ from our more complete and consistent model is accomplished with a modest variation in parameters from those detailed in \citet{ESPB2012}.

In addition to fitting \SNRJmm, we verify that our adaptation of the Blasi model matches the published work of the Blasi group, where appropriate, and show cases where our model differs from the implementation presented by Blasi and co-workers, detailing why these differences occur.
The most important difference is that we have coupled the updated steady-state model of Blasi and co-workers to the time-dependent, SNR hydrodynamics.\footnote{Some justification for using a steady-state DSA calculation in an 
evolving SNR is necessary. Basically, we assume that as long as the 
dynamic time-scale of the SNR is long compared to the acceleration time 
to $\pmax$, $\Tacc(\pmax)$, the steady-state approximation will be valid.
In all of the examples we show here, except for very
early times (i.e., $t\, \lsim 50$\,yr), we have
$\Tacc(\pmax) \ll \tSNR$.
We note that the quasi-steady-state model we use is also in good 
agreement with time-dependent models that show
``CR dilution"
\citep[see, for example][and references therein]{BEK1996a,BEK1996b,EB2011}.} 
Other differences concern the treatment of the CRs escaping at the free escape boundary (FEB) and how $\pmax$ is determined.

An alternative technique using a  \mc\ simulation of \NL\  (NL) DSA 
\citep*[e.g.,][]{JE91,VBE2009} makes a different set of assumptions from the \SA\ calculations and we have compared our results against this model as well. In this case, significant differences occur particularly concerning thermal particle injection. There are less pronounced differences in the maximum particle momentum, $\pmax$.

While our \crhydro\ code includes the influence of particle acceleration on the evolving SNR,   the  calculation of the 
thermal X-ray line emission coupled to the  CR production, and the production and propagation of escaping CRs to nearby mass concentrations, it remains a spherically symmetric model that doesn't attempt to describe the small-scale structure seen in most SNRs.
However, despite these approximations, we believe our generalized model is currently the most complete description of a young SNR in terms of the broad-band emission from radio to TeV
from a segment of the remnant large enough to safely average over the small-scale structure.
Future plans for the \crhydro\ code  include calculating the radiation emitted at the \RS\ and modeling the interaction of CRs with dense molecular clouds.

\section{Model}
\label{sec:model}
Supernova remnants are complex phenomena and any model with a claim to self-consistency between the remnant dynamics and the \NL\ production of CRs will be complicated. The time-dependent SNR hydrodynamics must be coupled to the CR production since the acceleration of CRs changes the ratio of specific heats for the shocked plasma and drains energy from the plasma as the highest energy CRs escape from the FS. The DSA of CRs is difficult to model  in a static configuration, particularly when the \NL\ shock structure, MFA,  and the \SC\ determination of the diffusion coefficient is included \citep*[e.g.,][]{VBE2008,BOE2011}. An evolving SNR adds additional complications \citep[e.g.,][]{BEK1996a,ZP2008,beopu2011}.

Given these complications, it is tempting to suggest that a 
particle-in-cell (PIC) simulation, rather than an analytic or \SA\ calculation, is the best approach 
\citep[e.g.,][]{RiqSpit2010}. In principle, a PIC simulation including all of the coupled phenomena could be done but in practice, for the immediate future at least, the dynamic range for particle energies and diffusion lengths preclude such a direct approach for the \nonrel\ shocks in SNRs
\citep[see][]{VBE2008}.
On the other hand, fully analytic self-similar \citep[e.g.,][]{ch82} or parameterized \citep[e.g.,][]{TM99,MC2011} approaches lack the ability to include \NL\ effects or MFA \SCly. Monte Carlo techniques have been used extensively to model \NL\ DSA including MFA 
\citep[e.g.,][]{VBE2008,VBE2009}. However, these techniques are computationally intensive and intrinsically steady-state and are not 
well-suited for inclusion in a SNR hydro simulation.
We believe the \SA\ approach developed here provides a compromise that 
is accurate enough to model observations of an evolving SNR and extract critical information on the supernova and environmental properties  until more complete numerical simulations are ``large" enough to fully model an evolving SNR.

When \rel\ CRs are produced efficiently
by the SNR shock, the  pressure to energy density ratio in the shocked plasma decreases.
In addition, some fraction of the highest energy CRs will escape from the SNR further reducing the pressure driving the FS. These effects result in a shock compression ratio larger than would be the case with \TP\ CR production and 
impact the SNR evolution. 
The \MFA\ further modifies the hydrodynamics by increasing the magnetic pressure downstream from the shock. 
Our model couples the hydrodynamics, DSA, and MFA allowing a \SC\ calculation of the CR spectra and resulting continuum radiation, as well as the thermal X-ray line emission in the interaction region between the FS and CD.
The changes produced in the hydrodynamics by NL DSA modify the ionization state of the shocked plasma and these changes are \SCly\ included in our \NEI\ calculation of the thermal X-ray emission.
The details of our generalized code are given below.

\subsection{Supernova remnant and shock hydrodynamics}
As in previous work \citep[e.g.,][]{ESPB2012}, the SNR is modeled with a spherically symmetric 
hydro simulation based on VH-1 \citep[e.g.,][]{BE2001}.\footnote{
The basic hydro simulation VH-1 is open to the public and can be found at {\tt http://wonka.physics.ncsu.edu/pub/VH-1/}. This 
web site includes a user guide and full background information on the code. We emphasize, however,  that while the hydrodynamics are essentially the same as in this basic code, the version we employ has been modified extensively to include NL DSA \citep[see][and references therein]{ESPB2012}.
} 
The standard hydrodynamic equations are modified when efficient DSA occurs through a change in the equation of state from the influence of \rel\ CRs, energy loss from escaping CRs, and magnetic pressure. The equation of state used in the hydro equations is modified by using the ratio of specific heats calculated from the particle distribution function so that the ratio lies between 5/3 and 4/3 depending on the fraction of \nonrel\ to \rel\ particle pressure.  
The SNR simulation provides the shock speed, sonic and \alf\ Mach numbers, and other quantities necessary for the shock acceleration calculation which is performed as follows.

From the conservation of momentum flux, we can write the following equation in the shock rest-frame:
\begin{eqnarray}
\rho_0 u_0^2 &+& P_{g,0} + P_{\mathrm{CR},0} + P_{w,0} = \nonumber \\
&& \rho(x)u(x)^2 + P_g(x) + P_\mathrm{CR}(x) + P_w(x)
\ .
\label{eqn:conserve}
\end{eqnarray}
Throughout this paper, we adopt a coordinate system in the shock rest-frame such that the subshock is located at $x=0$, and the subscripts `0', `1' and `2' refer to regions  far upstream ($x=-\infty$), immediately upstream ($x=0^-$) and immediately downstream ($x=0^+$) from the subshock, respectively. 
Equation~(\ref{eqn:conserve}) implicitly assumes that the shock is plane and we will always assume that the shock precursor is a small enough fraction of the shock radius for this to be a valid approximation 
\citep[see, for example,][for models that explicitly include the spherical symmetry and CR dilution]{BEK1996a,PZS2010}. Equation~(\ref{eqn:conserve}) further assumes that the shock obliquity is unimportant, i.e., that the  upstream magnetic field is nearly parallel to the shock normal. Essentially, we make the assumption that the magnetic field is tangled enough so that all of the effects of the magnetic field can be accounted for through a scalar magnetic pressure.
In equation~(\ref{eqn:conserve}),  
$\rho_0$ is the far upstream density including helium if it is present,
$\gamma_g = 5/3$ is the ratio of specific heats for the thermal gas, $M_0$ is the sonic Mach number,
$P_{g,0} = {\rho_0 u_0^2}/(\gamma_g M_0)$ is the ambient gas pressure, 
$P_{\mathrm{CR},0}$ is the pressure from any pre-existing seed CR proton population, 
$P_\mrm{CR}(x)$ includes the pressure from the accelerated protons,
$P_{w,0}$ is the pressure from any pre-existing ambient magnetic turbulence,
and $P_w(x)$ includes the pressure from the 
self-generated magnetic turbulence through resonant CR streaming instability.

We include precursor heating from the damping of turbulent waves and adopt equation (50) from 
\citet{BE99}, such that the gas pressure at any $x<0$ in the precursor region is given by
\begin{eqnarray}
P_g(x) \cong && P_{g,0}U(x)^{-\gamma_g} \times \nonumber \\
&& \left[1+\fDamp(\gamma_g-1)\frac{M_0^2}{M_{A,\mathrm{eff}}}(1-U(x)^{\gamma_g})\right] ,
\label{preheating}
\end{eqnarray} 
where $U(x) \equiv u(x)/u_0$ and $M_{A,\mathrm{eff}} \equiv \sqrt{M_{A,0} M_{A,1}} = \sqrt{u_0u_1/(v_{A,0} v_{A,1})}$.\footnote{Note that in many formulations $M_{A,0}$ is assumed in equation~(\ref{preheating}) rather than $M_{A,\mathrm{eff}}$ but for most cases where the \alf\ Mach is large, this is not significant.} 
Here we have assumed that the 
speed of the Alfv\'{e}n waves $v_A$ is unimportant compared to the shock speed $u_0$, which is always the case for the models described in this paper. 
A model of wave-damping such that the wave-growth rate $\sigma(x)$ is proportional to the damping rate $\Gamma(x)$ is used, so that $\Gamma(x) = \fDamp \sigma(x)$, where $\fDamp \in [0,1]$ is a free parameter of the model.

For the magnetic pressure, we refer to the quasi-linear model in \citet{CAB2010} and express the magnetic pressure in the precursor region in terms of $U(x)$ as
\begin{eqnarray}
P_w(x) &=& \frac{\delta B(x)^2}{8\pi} \nonumber \\
           &=& \frac{\delta B_\mathrm{ISM}^2}{8\pi} + \frac{(1-\fDamp)\rho_0u_0^2}{4M_{A,0}}\left(\frac{1-U(x)^2}{U(x)^{3/2}}\right)
\ ,
\label{Pw}
\end{eqnarray}
where $\delta B_\mathrm{ISM}$ is the background turbulent field strength in the far upstream medium. For simplicity, we do not consider the presence of pre-existing magnetic turbulence in the ISM and assume $\delta B_\mathrm{ISM} = 0$ (or $P_{w,0} = 0$) for all models presented in this paper. The total amplified $B$-field in the precursor is hence 
\begin{eqnarray}
B(x) &=& \sqrt{\delta B(x)^2 + B_0^2}  \nonumber \\
       &=& \sqrt{\frac{8\pi(1-\fDamp)\rho_0u_0^2}{4M_{A,0}}\left(\frac{1-U(x)^2}{U(x)^{3/2}}\right) + B_0^2}
\ ,
\label{totalB}
\end{eqnarray}
since we assume that the generated turbulence is purely Alfv\'{e}nic and transverse to the background field $B_0$.  
For models in which MFA is not included, $P_w(x)$ is set to zero everywhere.

The CR pressure $P_\mathrm{CR}$ is given by 
\begin{eqnarray}
P_{\mathrm{CR}}(x) = && P_{\mathrm{CR},0}U(x)^{-\gamma_\mathrm{CR}(x)} + \nonumber \\
                                        && \frac{4\pi}{3}\int_{p_\mathrm{inj}}^{p_\mathrm{max}}p^3v(p)f(x,p)dp 
\ ,
\end{eqnarray} 
where $f(x,p)$ is the phase-space distribution of the accelerated particles and $\gamma_\mathrm{CR}(x) = 1 + P_\mathrm{CR}(x)/E_\mathrm{CR}(x)$ is the ratio of specific heats for the CR particles. 
The production of $f(x,p)$ via \NL\ DSA is described in 
Section~\ref{Sec:DSA}.
Finally, we can rearrange equation~(\ref{eqn:conserve}) in a dimensionless form by dividing it with $\rho_0 u_0^2$:
\begin{eqnarray}
\crPup(x) = && 1 - U(x) + \nonumber \\
& & \left(\frac{1}{\gamma_g M_0^2} +\frac{\fDamp(\gamma_g-1)}{\gamma_gM_{A,\mathrm{eff}}} \right) \left( 1-U(x)^{-\gamma_g} \right) + \nonumber \\
& & \frac{\crPgas}{\gamma_g M_0^2}\left( 1-U(x)^{-\gamma_\mathrm{CR}} \right) + \nonumber \\
& & \frac{1-\fDamp}{4M_{A,0}}\left( \frac{1-U(x)^2}{U(x)^{3/2}}\right)
\ ,
\end{eqnarray}
where $\crPup(x) \equiv P_\mathrm{CR}(x)/(\rho_0u_0^2)$ and 
$\crPgas \equiv P_{\mathrm{CR},0}/P_{g,0}$. 
The third term on the right-hand-side (RHS) describes the adiabatic compression and wave-heating in the precursor, the fourth term on the RHS describes the pre-shock compression of the incoming pre-existing non-thermal
particles, and the last term is the pressure from the self-generated turbulence.

Following \citet{CAB2010}, the total and subshock compression ratios experienced by the background fluid, $R_\mathrm{tot}$ and $R_\mathrm{sub}$ respectively, are related through the following equation:
\begin{eqnarray}
R_\mathrm{tot}^{\gamma_g+1} &=& \frac{M_0^2R_\mathrm{sub}^{\gamma_g}}{2}\left[\frac{(\gamma_g+1)-R_\mathrm{sub}(\gamma_g-1)}{(1+\Lambda_B)(1+\Lambda_\mathrm{TH})}\right] \nonumber \ , \\
\Lambda_B &\equiv& \frac{P_{w,1}}{P_{g,1}}\left[1 + R_\mathrm{sub}\left(\frac{2}{\gamma_g}-1\right)\right] \nonumber \ , \\
\Lambda_\mathrm{TH} &\equiv& \fDamp(\gamma_g-1)\frac{M_0^2}{M_{A,\mathrm{eff}}}(1-U_1^{\gamma_g}) 
\ ,
\end{eqnarray} 
where $\Lambda_B$ and $\Lambda_\mathrm{TH}$ give the corrections from the magnetic jump condition with self-generated turbulence, and from precursor heating, respectively. Equipped with the compression ratios, one can readily write down the gas temperatures in the upstream precursor $T(x)$ and immediately downstream from the shock, $T_2$:
\begin{eqnarray}
T(x) &=& T_0 \left[\frac{P_{g}(x)}{P_{g,0}}\right]U(x) \nonumber \\
       &=& T_0\  U(x)^{1-\gamma_g} \times \nonumber \\
               & & \left[1+\fDamp(\gamma_g-1)
               \frac{M_0^2}{M_{A,\mathrm{eff}}}(1-U(x)^{\gamma_g})\right]
\ ,
\end{eqnarray}
and
\begin{eqnarray}
T_2 &=& T_1  \left(\frac{P_{g,2}}{P_{g,1}}\right)R_\mathrm{sub}^{-1} \nonumber \\
       &=& \textstyle T_1 \left[\frac{(\gamma_g+1)R_\mathrm{sub} 
                   - (\gamma_g -1) + (\gamma_g-1)(R_\mathrm{sub}-1) \Delta}
                      {(\gamma_g+1-(\gamma_g-1)R_\mathrm{sub})R_\mathrm{sub}}\right] \nonumber \ , \\
\Delta &\equiv&  \frac{P_{w,1}}{P_{g,1}} (R_\mathrm{sub}-1)^2 \nonumber \\
          &=& \textstyle \frac{\gamma_gM_0^2 (1-\fDamp) (1-U_1^2) (R_\mathrm{sub}-1)^2}
                          {4M_{A,0}\left[1+\fDamp(\gamma_g-1) (M_0^2/M_{A,\mathrm{eff}}) (1-U_1^{\gamma_g})\right]
                          U_1^{3/2-\gamma_g}} \ . \nonumber \\
\end{eqnarray}

\begin{figure}
\centering
\includegraphics[width=8cm]{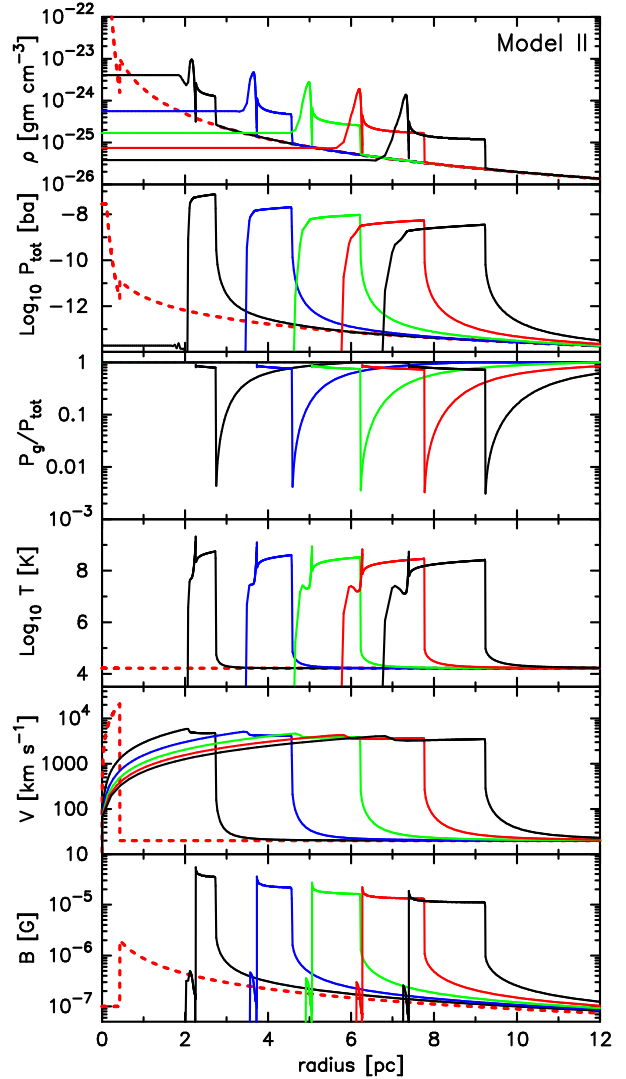} 
\caption{Lab-frame snapshots of the density ($\rho$), total pressure ($\Ptot \equiv P_\mrm{CR} + P_g$), gas to total pressure ratio ($P_g/\Ptot$), temperature ($T$), flow speed ($V$), and magnetic field ($B$) as a function of radius from the explosion center at equally spaced  times after the explosion. The parameters used are those  of Model \Rmnum{2} in  
Table~\ref{table:param}. The red dotted lines show the initial conditions of the model at $t_0=20$\,yr. 
The parameter for the FEB is $\fFEB=0.2$  and the precursor is clearly evident in all of the profiles except for the density. The density precursor is present but is much less pronounced as explained in the text. The CD shows up as a sharp density drop in the top panel and the RS is present but no attempt has been done to accurately resolve it.
The behavior of $P_g/\Ptot$ comes about because the thermal pressure in the pre-SN wind is very low due to the low temperature and small gas density which drops with radius, such that $\Ptot$ is always dominated by the CR pressure $P_\mrm{CR}$ in the precursor. At the subshock, $P_g/\Ptot$ jumps from $\sim 0.003$ to $\sim 0.7$ mainly due to the strong shock heating of the upstream gas to $T \gtsim 10^8$\,K, and becomes unity behind the CD since particle acceleration is not considered on the reverse shock side in this work.} 
\label{fig:precursor}
\end{figure}

We note that the hydro code now includes the CR precursor \SCly\ producing a smooth transition from the far upstream values of density, pressure, temperature, flow speed, and $B$-field  to the subshock values. This has been included for consistency but does not significantly modify the hydrodynamics because the precursor drops sharply upstream from the subshock, as shown in 
Figure~\ref{fig:precursor}.\footnote{The lab-frame density precursor is present in Figure~\ref{fig:precursor} but it is much less pronounced than the others. The reason for this can be seen by comparing $\rho(x)$ and $V(x)$. The density is given by  $\rho(x)/\rho_0 =1/U(x)$ and when the shock is modified,  
$U(x)<1$ and the incoming fluid is slowed by a fraction $1-U(x)$. When the constant wind is taken into account, the lab-frame plasma speed is
$V(x)=\VelFS(x) - U(x) (\VelFS - \Vwind)$. In this particular example using parameters that fit \SNRJmm, $U(x)$ ranges from 0.7 to 1.0 at most, accounting for the small variation in $\rho(x)$. The plasma speed, however, is 
$V(x)/\Vwind = [1 - U(x)] (\VelFS/\Vwind) + U(x)$
and $\VelFS/\Vwind \sim 200-300$, making the speed precursor much more prominent.}
However, the presence of a precursor is an important prediction and diagnostic for NL DSA and our calculation will be important for future determinations of the precursor emission.\footnote{The sharp density and temperature features at the CD are artifacts of the initial conditions for the hydro simulation (see the feature where the ejecta joins the ISM material in the dashed red curve in the top panel of 
Figure~\ref{fig:precursor}) and the assumption of spherical symmetry. Similar features are present in self-similar solutions \citep[e.g.,][]{ch82}. The amount of material in the sharp density region is generally small compared to the total in the shocked \CSM\ and does not bias the total thermal emission in any noticeable way.}
%

\subsection{Diffusive Shock Acceleration}
\label{Sec:DSA}
The acceleration of thermal particles, or the (re)acceleration of ambient Galactic CRs, to relativistic energies through DSA can be described by the following diffusion-advection equation written in the shock rest-frame,
\begin{eqnarray}
&& \left[u(x)-v_A(x)\right]\frac{\partial{f}}{\partial{x}} - Q(x,p)\nonumber \\
&=& \frac{\partial}{\partial{x}}\left[D(x,p)\frac{\partial{f}}{\partial{x}}\right] + \frac{d[u(x)-v_A(x)]}{dx}\frac{p}{3}\frac{\partial{f}}{\partial{p}} 
\ ,
\label{eq:diffeq}
\end{eqnarray}
where the diffusion coefficient $D(x,p)$ depends both on the momentum of the particles and the position in the precursor, and $Q(x,p)$ is the injection rate from 
the shock-heated plasma.
The intrinsic assumptions that the phase-space distribution of the accelerated particles $f(x,p)$ is nearly isotropic in the fluid frame, and that the shock is non-relativistic are used in this equation. 
The diffusion coefficient should vary with $x$ as the local $B$-field changes due to MFA in the CR precursor. 
As an option, we can parameterize $D(x,p)$ in the following form: 
\begin{equation}
D(x,p) = D_0\left[\frac{3\mu \mrm{G}}{B(x)}\right]\beta\left(\frac{p}{p_0}\right)^\alpha
\ ,
\label{eq:diff}
\end{equation}
where $p_0 = 1$~GeV$/$c unless otherwise specified, 
$\beta \equiv v(p)/c$ with $v(p)$ being the particle velocity, and $D_0$ is a normalization factor. Here, $B(x)$ is the amplified magnetic field in the precursor region. For Bohm diffusion, for example, we have $\alpha = 1$ and $D_0 = p_0 c^2/[3e(3 \mu \mrm{G})]$.

Alternatively, we can compute the diffusion coefficient by extending a quasi-linear theory of wave generation \citep[e.g.,][]{BE87} to the high-amplitude regime as a first approximation. In this case, $D(x,p)$ is calculated from the self-generated wave spectrum with a power-law form $W(x,k) = W_0(x)(k_0/k)^{2-\alpha}$, where $k$ is the wavenumber and is related to the particle momentum through gyro-resonance, i.e., $k_\mathrm{res} = e\delta B(x)/(pc)$. Note that this resonance relation is space dependent due to MFA in the CR precursor.
The normalization factor $W_0(x)$ can be obtained 
self-consistently using the pressure from magnetic turbulence $P_w$ through equation~(\ref{Pw}) and the relation $P_w(x) = \frac{1}{2}\int W(x,k)dk$ for Alfv\'{e}n waves. The resonant scattering mean-free-path is given by $\lambda_\mrm{res} = (4/\pi ) P_w(x)/[k^2 W(x,k)]$. Given these assumptions, we can readily obtain the diffusion coefficient as $\lambda_\mrm{res}v / 3$,
\begin{eqnarray}
D(x,p) = \begin{cases} \frac{2pc^2\beta}{3\pi e\delta B(x)}\mrm{ln}\left(\frac{\pmax}{p_\mrm{inj}}\right),  & \mbox{ if }  \alpha = 1  \\ 
\frac{2\pmax c^2 \beta}{3\pi (1-\alpha) e \delta B(x)}  \left(\frac{p}{\pmax}\right)^\alpha, & 
\mbox{ if } \alpha < 1 \end{cases}
\ .
\label{Dxp}
\end{eqnarray}

It can be shown that the distribution of accelerated particles at the subshock can be well approximated by the following solution 
\citep[e.g.,][]{B2004, AB2005, CAB2010}: 
\begin{eqnarray}
&& \fzCR = \frac{3S_\mathrm{tot}}{S_\mathrm{tot}U(p)-1} \times \nonumber \\ 
& & \Bigg\lbrace \frac{\eta n_0}{4\pi p_\mathrm{inj}^3}
\mathrm{exp} \left[ -\int_{p_\mathrm{inj}}^p
 \frac{dp'}{p'} \frac{3S_\mathrm{tot}U(p')}{S_\mathrm{tot}U(p')-1} \right] + \nonumber \\
& & \int_{p_\mathrm{min}}^p \frac{dp''}{p''} f_\infty(p'') 
\mathrm{exp} \left[ -\int_{p''}^p \frac{dp'}{p'} 
\frac{3S_\mathrm{tot}U(p')}{S_\mathrm{tot}U(p')-1} \right] \Bigg\rbrace \ , \nonumber \\
\label{eqn:f0}
\end{eqnarray}
Here we have adopted the thermal injection model described in e.g.~\citet{BGV2005}, such that  
$Q(x,p) = [\eta n_1 u_1/(4\pi p_\mrm{inj}^2)]\delta (x) \delta(p-p_\mrm{inj})$ is the injection rate,
$\eta \equiv \frac{4}{3\sqrt{\pi}} (S_\mathrm{sub}-1)\Inj^3 e^{-\Inj^2}$ is the fraction of thermal particles being injected into DSA, and $\Inj$ is defined such that the injection momentum in the shock frame is $p_\mathrm{inj} = (\Inj - u_2/c) \sqrt{2m_pkT_2}$ 
\citep[e.g.,][]{CKVJ2010}. The `cr' superscript on $\fzCR$ indicates that only superthermal particles are included in 
equation~(\ref{eqn:f0}). 
The first term in the braces in 
equation~(\ref{eqn:f0}) describes the acceleration of the injected particles from the downstream thermal pool, while the second term describes the acceleration of the pre-existing CRs in the ISM, with an ambient spectrum $f_\infty(p)$. 
Here, for simplicity, we have assumed that the minimum momentum $p_\mathrm{min} = p_\mathrm{inj}$ for the pre-existing CRs. 
To produce a full spectrum of all shocked particles, $f_1(p)$, we  add a thermal component to $\fzCR$ with a density and temperature that are consistent with the NL DSA calculation, producing a spectrum similar to the red solid curves shown in 
Figure~\ref{fig:static_test}.

The `effective' compression ratios $S_\mathrm{tot}$ and $S_\mathrm{sub}$ experienced by the streaming particles scattering off the magnetic turbulence are smaller than those felt by the background fluid if the scattering centers are moving away from the shock upstream in the plasma frame. We assume that the magnetic scattering centers have a local speed, relative to the plasma, of
%
\begin{eqnarray}
v_A(x) &=& \frac{B_0} {\sqrt{4\pi\rho(x)}}\Bigg\lbrace 1 + \left[\frac{B(x)}{B_0} - 1\right] \falf \Bigg\rbrace \nonumber \\
             &=& v_{A,0}\sqrt{U(x)} \Bigg\lbrace 1 + \left[\frac{B(x)}{B_0} - 1\right] f_\mrm{alf} \Bigg\rbrace
\label{eq:vAlf}
\end{eqnarray}
in the direction parallel to the shock normal and away from the shock. The model parameter $\falf \in [0,1]$, allows us to vary $v_A(x)$ linearly
between $B_0/ \sqrt{4\pi\rho(x)}$ and $B(x)/\sqrt{4\pi\rho(x)}$. 
When $\falf = 0$, $v_A$ corresponds exactly to the definition of the Alfv\'{e}n speed, which is believed to represent well the velocity of the magnetic turbulence in the limit of small-amplitude $\delta B  \ll B_0$. As the self-generated turbulence deviates from the small-amplitude regime, however, the wave speed may also deviate from the Alfv\'{e}n speed. We use the linear parametrization in equation~(\ref{eq:vAlf}) as a simple model to reflect this transition. Similar approaches have been used previously by several other groups \citep[e.g.][]{BL2001,VBE2008,ZP2008,CBAV2008}. We note that this recipe can be modified as theories and/or simulations will improve.
With this definition of $v_A(x)$, we can obtain:
\begin{eqnarray}
S_\mathrm{tot} &=& \frac{u_0-v_{A,0}}{u_2+v_{A,2}} \nonumber \\
                         &=& R_\mathrm{tot}  \left(\frac{1 - V_{A,0}}
                                 {1+V_{A,0}[1+(B_2/B_0-1)f_\mrm{alf}]\sqrt{R_\mathrm{tot}}}\right) \nonumber \ , \\
S_\mathrm{sub} &=& \frac{u_1-v_{A,1}}{u_2+v_{A,2}} \nonumber \\
                          &=& R_\mathrm{sub}  \left(\frac{1 - V_{A,0}[1+(B_1/B_0-1)f_\mrm{alf}]U_1^{-1/2}}
                                 {1+V_{A,0}[1+(B_2/B_0-1)f_\mrm{alf}]\sqrt{R_\mathrm{tot}}}\right) \ , \nonumber \\
\end{eqnarray}
where $V_A \equiv v_A/u_0$ is the dimensionless wave speed. 
The general effect of using the amplified magnetic field instead of the background field in equation~(\ref{eq:vAlf}) is a larger wave velocity and a decrease in the effective compression ratios, and hence a less modified shock structure and reduced acceleration efficiency. This effect is especially notable for cases when MFA is efficient.
The quantity $U(p)$ in equation~(\ref{eqn:f0}) roughly indicates the flow speed experienced by a particle with 
momentum $p$ at one diffusion length upstream from the subshock, 
and is given by:
\begin{eqnarray}
U(p) &=& \int_{-\infty}^{0} dx \frac{[U(x) - V_A(x)]^2}{x_p(x,p)} \times \nonumber \\
            && \mathrm{exp}\left[ -\int_{x}^{0} dx'  \frac{U(x')-V_A(x')}{x_p(x',p)}\right]
\ ,
\end{eqnarray}
with 
$x_p(x,p) \equiv 3S_\mathrm{sub}D(x,p)/[(S_\mathrm{sub}-1)q(p)u_0]$ 
and 
$q(p) \equiv - \partial{\mathrm{ln}[\fzCR]}/\partial{\mathrm{ln}(p)}$.

To determine the field compression at the subshock, we follow  previous work \citep[e.g.,][]{Reynolds98} and assume that the 
$B$-field is totally random in orientation. The downstream field is then given by  $B_2 = B_1\sqrt{1/3 + 2/3R_\mathrm{sub}^2}$, where $B_1$ is the field strength immediately upstream of the subshock. 

When the above elements are combined, we obtain the superthermal particle distribution in the precursor at any position $x$:
\begin{equation}
f^\mathrm{cr}(x,p) = f^\mathrm{cr}_1(p) \mathrm{exp}\left[ - \int_{x}^{0} dx' \frac{U(x') - V_A(x')}{x_p(x',p)} \right]
\ ,
\end{equation}
To obtain the full spectrum, $f(x,p)$, a thermal portion has been added at each $x$ with a density and temperature consistent with the precursor values at that position in a similar way to $f^\mathrm{cr}_1(p)$ at the subshock.

\subsubsection{Maximum particle momenta}
\label{sec:pmax}
The determination of the maximum momenta of the accelerated protons and electrons is done self-consistently using the precursor information and diffusion coefficient. 
For protons, $p_\mathrm{max}$ is obtained by either equating the acceleration time-scale with the remnant age (age-limited), or by equating the upstream
diffusion length with a `free escape boundary' (escape-limited) set in front of the shock at a distance $L_\mathrm{feb} \equiv f_\mathrm{feb} R_\mathrm{SNR}(t)$, where $\fFEB$ is typically set between 0.1 and 0.2.
Whichever limit gives the lowest $\pmax$ is used. 
The acceleration time-scale is approximated as 
\citep[e.g.,][]{Drury83,BE99}: 
\begin{equation}
t_\mrm{acc} \approx \frac{3}{u_0-u_2} \int_{p_\mrm{inj}}^{p_\mrm{max}} \frac{dp}{p} \left(\frac{D_0(p)}{u_0} + \frac{D_2(p)}{u_2}\right)
\ .
\end{equation}
To obtain the escape-limited $\pmax$, we equate a diffusion length averaged over the precursor,
i.e.,
\begin{equation} 
\langle D(x,\pmax)/u(x) \rangle = \int_{-L_\mathrm{feb}}^{0} [D(x,\pmax)/u(x)] dx/L_\mathrm{feb}
\ ,
\end{equation}
with the free escape boundary distance. 
Note that in the case when the diffusion coefficient is determined from the self-generated wave spectrum, $D(x,p)$ is dependent on $\pmax$ as
shown in equation~(\ref{Dxp}). Therefore, $\pmax$ has to be obtained through recursive steps in the code.  

For electrons, $p_\mathrm{max}$ can also be limited by radiation losses, mainly from \syn\ and inverse-Compton losses. The loss time scale of a relativistic electron, $t_\mrm{loss} = \gamma/\dot{\gamma}$,  is given by  
\citep[i.e.,][]{SchlickeiserEtal2010}:
\begin{equation}
t_\mrm{loss} = \frac{3 m_e c^2}{4\sigma_T c U_B \gamma}\left( 1 + \sum_{i=1}^{N_\mrm{phot}} \frac{W_i \gamma_{k,i}^2}{U_B(\gamma^2+\gamma_{k,i}^2)} \right)^{-1}
\ ,
\end{equation}
where we define $U_B \approx B_2^2/8\pi$, $\gamma$ is the Lorentz factor of the electrons, $\Nphot$ is the number of seed photon fields, and $\gamma_{k,i} = 0.53m_ec^2/k_BT$ and $W_i$ are the critical 
Klein-Nishina Lorentz factor and energy density of the $i$-th seed photon field, respectively.
For simplicity, only the CMB photon field is considered for the models in this paper. We then obtain the loss-limited $p_\mathrm{max}$ by equating the loss time-scale with the acceleration time-scale. 
When radiation  loss is slow relative to acceleration, the $\pmax$ of electrons is the same as for protons.

\subsubsection{Escaping cosmic rays}
The final process that needs to be accounted for to obtain a \SC\ solution for the CR production is the energy flux of particles that escape upstream from the shock at the FEB.
When $\pmax$ is set by escape, the highest energy CRs leave the FEB with a flux determined by energy conservation and is given by:
\begin{eqnarray}
&& F_\mathrm{esc} =
                             \left(\frac{1}{2}\rho_0 u_0^3 + \frac{\gamma_g}{\gamma_g-1}P_{g,0}u_0 + 
                             \frac{\gamma_\mathrm{CR,0}}{\gamma_\mathrm{CR,0}-1}P_{\mathrm{CR},0}u_0\right) -
                             \nonumber \\
                            && {} \left(\frac{1}{2}\rho_2 u_2^3 + \frac{\gamma_g}{\gamma_g-1}P_{g,2}u_2 +
                             \frac{\gamma_\mathrm{CR,2}}{\gamma_\mathrm{CR,2}-1}P_\mathrm{CR,2}u_2 +
                             F_{w,2}\right) \ . \nonumber \\
\label{eqn:Fesc}
\end{eqnarray}
In equation~(\ref{eqn:Fesc}), the magnetic energy flux downstream from the shock, in the limit $v_{A,2} \ll u_2$, can be approximated as 
$F_{w,2} \approx 3 u_2 P_{w,2} = 3 u_2 P_{w,1} (1/3+2/3R_\mathrm{sub}^2)$.
The shape of the escaping CR distribution is determined as in \citet{EB2011}.

Once the above equations are formulated, the NL DSA problem is solved by  making an initial guess for $U_1$ and then using this to calculate the quantities $U(x)$,  $\crPup(x)$ and hence $U(p)$,  
$B(x)$, $D(x,p)$, $p_\mathrm{max}$ and $f_1(p)$. 
The \SC\ result, where all of the above quantities are  mutually consistent and the system of equations is internally satisfied, is  determined through iteration using the scheme described in 
\citet{CAB2010}.

\subsubsection{NEI and temperature equilibration}
\label{sec:TempEq}
%
%
%
%

Typical SNRs have densities that are low enough that the ionizaton state of the shocked gas can differ significantly
from equilibrium. The approach to equilibrium depends on the electron and ion densities ($n_e$, $n_i$), the electron temperature ($T_e$), 
and the time for which the shocked gas has been ionizing. Since these quantities are also affected by the diffusive
shock acceleration process \citep[e.g.,][]{EPSBG2007}, we have incorporated a calculation into the model that follows the
nonequilibrium ionization of the shocked gas \citep*[i.e.,][]{PES2009}, and couples the resultant ionization vectors, 
densities, and electron temperatures to a plasma emissivity code \citep[i.e.,][]{PSRE2010}.  The ionization structure of the shock-heated gas at a particular distance behind the shock in a SNR is determined by the electron density $n_e$, the electron temperature $T_e$, and the ionization and recombination rates for each ion of interest. This structure is calculated by solving the collisional ionization equations in a Lagrangian gas element behind the shock as described in detail in \citet{PES2009}.

Since the hydro simulation does not specify the individual electron and proton temperatures, a temperature equilibration model for the postshock region must be assumed. Two extreme cases for electron heating are clear:  instant equilibration
with the proton temperature, and equilibration through Coulomb collisions \citep[i.e., equation 5-30 in][]{spitzer62}. For both cases we include the effects of
adiabatic expansion on the temperature components. 
While all of the examples shown in this paper assume that the electrons
are heated by Coulomb collisions, the \crhydro\ code allows for a continuous range of heating modes between the two extremes. Examples with faster heating modes are shown in \citet{EPSBG2007} and \citet{PES2009}. 

\subsubsection{Effect of ion-neutral damping in a partially-ionized medium}
It is known that Alfv\'{e}n wave evanescence due to ion-neutral collisions in a partially-ionized precursor can lead to wave damping and hence a reduction of trapping power for the highest energy particles. Recent gamma-ray observations by \textit{Fermi}, combined with previous TeV observations by ground-based Cherenkov telescopes, have revealed high-luminosity, broken power-law gamma-ray spectra 
from nearly a dozen Galactic SNRs which are known to be interacting directly with adjacent molecular clouds 
\citep[e.g.][]{AbdoEtalW51C2009,AbdoEtalW442010,AbdoEtalIC4432010}. It has been postulated immediately after the discovery that ion-neutral damping can naturally give rise to a spectral break in the accelerated particles for fast molecular shocks, which may provide an explanation for these peculiar gamma-ray spectra 
\citep[e.g.,][]{MDS2011,uchiyamaea10}. 

In order to model shocks running into a partially-ionized material, 
we have included phenomenological features 
following previous studies by \citet{DDK96, bceu00, MDS2011} and references therein.
The presence of a spectral break in momentum, $p_\mathrm{br}$, originates from the fact that particles with $p > p_\mathrm{br}$ are only weakly scattered by non-resonant small-scale waves. The diffusion coefficient is hence modified in the precursor region due to a transition of the wave spectrum below and above $p_\mathrm{br}$. 
When ion-neutral damping is expected to be important, we follow the parameterization in \citet{bceu00} and impose a momentum break for the diffusion coefficient at $p_\mathrm{br}$ such that $D \propto (p/p_\mathrm{br})^\alpha$ for $p \le p_\mathrm{br}$ and $D \propto (p/p_\mathrm{br})^2$ for $p > p_\mathrm{br}$, respectively.
The momentum $p_\mathrm{br}$ can be estimated by equating the gyro-radius to the reciprocal of the evanescence wavenumber 
$\Gamma _\mathrm{in} = 2\pi v_a / \nu _\mathrm{in}$, where $\nu _\mathrm{in}$ is the ion-neutral collision frequency. Following \citet{MDS2011} and references therein, we write:
\begin{equation}
p_\mathrm{br} = 10 (T_4^{-0.4} n_0^{-1} n_i^{-0.5} B_{\mu}^2)\ m_p c 
\ ,
\end{equation}
where the upstream quantities, $n_{0,i}$, are the number densities of neutral and ionized particles respectively, $T_4 = T/10^4$~K, and $B_{\mu} = B/1\ \mu$G. Assuming that the pitch-angle distribution of the particles below the break momentum remains locally isotropic, and that the underlying spectrum extends to arbitrarily high energy, the spectrum steepens by one power-law index at $p_\mathrm{br}$, i.e., $f(p) \rightarrow f(p)(p_\mathrm{br}/p)$ for $p \ge p_\mathrm{br}$. 

In addition, wave-damping due to ion-neutral collisions can result in a high-energy cut-off by imposing limits on the maximum energy achievable by the accelerated particles through DSA. 
We follow \citet*{DDK96}, who estimates this maximum energy to be
\begin{equation}
E_\mathrm{max,in} = u_{0,7}^3 T_4^{-0.4} n_n^{-1} n_i^{0.5} 
\xi_\mathrm{CR,-1}\ \mathrm{GeV}
\ ,
\end{equation}
where $u_{0,7} = u_0/10^3$\,\kmps, and  
$\xi_\mathrm{CR,-1} = \crPup/0.1$ evaluated at the subshock. This puts an extra constraint on $p_\mathrm{max}$ in addition to the factors described in the previous section, when the upstream medium is only partially ionized. We reserve a detailed discussion of 
ion-neutral damping for future work. 

\subsubsection{Cases with a pre-SN wind}
In the above equations it was implicitly assumed that the shock was propagating into a uniform medium as might be the case for thermonuclear, Type Ia supernovae. For core-collapse SNe, it is expected that pre-SN winds will  produce spatial gradients for the
pre-shock quantities like the background $B$-field, gas density, and gas pressure, which are dependent on the properties of the wind such as the mass loss rate of the progenitor star. These gradients have to be taken into account when the recipe described in the previous sections is applied, especially when we calculate the profiles of various quantities in the CR precursor.  
In these cases, we modify, for example, 
equations~(\ref{preheating}) and (\ref{totalB}) by replacing the constant background values $P_{g,0}$ and $B_0$ with the 
space-dependent values in the pre-shock wind, and similarly for the gas density and velocity. Throughout this paper, we assume an isothermal and constant velocity wind whenever one is applied.


\begin{figure}
\centering
\includegraphics[width=8cm]{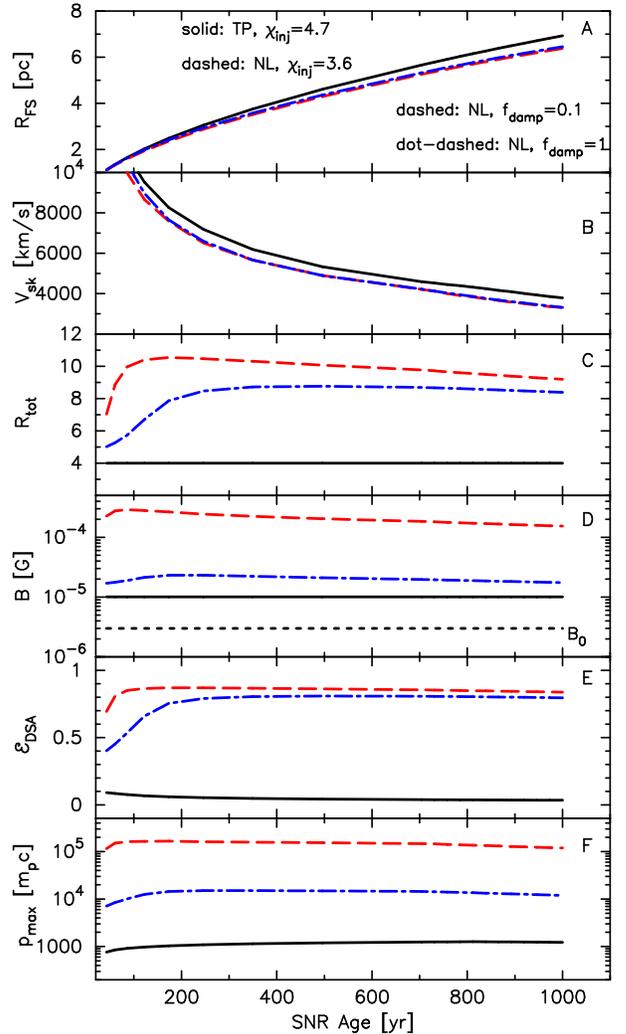} 
\caption{Evolution of a SNR with \TP\ DSA ($\Inj=4.7$; solid, black curves, Model III) and efficient DSA [$\Inj=3.6$; dashed (red) curves Model IV, and dot-dashed 
(blue) curves Model V]. Panel A shows the FS radius, panel B shows the FS speed, panel C shows the FS total compression ratio, panel D shows the magnetic field, $B_2$, immediately downstream (DS) from the FS 
(the curve marked $B_0$ is the unshocked field for all three models),  panel E shows the DSA efficiency at any given time, and panel F shows the proton $\pmax$. These models all have a uniform ISM with $\EnSN=10^{51}$\,erg, $\Mej=1.4\,\Msun$, 
 $n_0=0.1$\,\pcc, $B_0=3$\,\muG\ (shown as a dotted line in panel D), $T_0=10^4$\,K, $\tSNR=1000$\,yr, and $\fAlf=0$. The starting time for the simulation is $t_0=30$\,yr and the transition from 
age-limited to size-limited acceleration occurs smoothly with no noticeable change in the SNR evolution at $t\, \lsim  40$\,yr.} 
\label{fig:TPvsNL}
\end{figure}

\section{Results}
We first show in Section~\ref{sec:TPvsNL} how efficient DSA can modify the evolution of a SNR with a simple example of a SNR exploding in a homogeneous environment typical of Type Ia supernovae.
Then, since the
generalizations described above closely follow the work of Blasi and co-workers, we compare our results to recent work from that group both with and without MFA in 
Section~\ref{sec:CapEtal}. 
It is important to note that alternative descriptions of NL DSA exist and in 
Section~\ref{sec:CapEtal} we compare our results to the \mc\ calculations of \citet{EV2008} and \citet{CKVJ2010}.
In 
Section~\ref{sec:J1713} we give a detailed discussion of \SNRJ\ and compare our results against those of \citet{ESPB2012}.  

\subsection{Modified evolution of SNRs with self-generated magnetic turbulence} 
\label{sec:TPvsNL}
In this section we compare the time evolution of a SNR undergoing efficient DSA with MFA to a  \TP\ {(TP) case. 
The solid (black) curves in
Figure~\ref{fig:TPvsNL} are for the TP example (Model III, $\Inj=4.7$) and the dashed (red) and dot-dashed (blue) curves have efficient acceleration 
($\Inj=3.6$). For the dashed (red) curves, the wave damping factor $\fDamp=0.1$ (Model IV), while for the dot-dashed (blue) curves $\fDamp=0.99$, suppressing MFA (Model V).
As expected, efficient DSA results in a slower FS speed and a smaller SNR radius at a given age than for the TP case. The proton $\pmax$ is increased dramatically for the NL case and is $>100$ times the TP $\pmax$ for $\fDamp=0.1$.
When $\fDamp$ approaches $1$ and MFA is suppressed, the changes between the TP and NL results are less pronounced but the proton $\pmax$ is still $\sim 10$ times greater in the NL case.
As shown in Table~\ref{table:ModelOut}, we note that the $\pmax$ of electrons is limited by fast synchrotron cooling for Model IV due to the highly amplified downstream B-field $B_2$ from efficient MFA. On the other hand, Model III and V produce accelerated electrons with $\pmax$'s essentially the same as the protons. In these cases, MFA is inefficient due to a low DSA efficiency (TP) and a high wave damping rate respectively, and the resulted $B_2$ is not large enough to have any importance on determining the electron cutoff's.
Also, in this example, since $B_0=3$\,\muG\ is relatively small, the magnetic field has little influence on the SNR evolution as indicated by the fact that the dashed and dot-dashed curves for $\RFS$ and $\VelFS$ are almost indistinguishable. 

\subsection{Comparison with \citet{CAB2010,CKVJ2010}}
\label{sec:CapEtal}
In the top panel of Figure~\ref{fig:static_test}, we show a proton spectrum presented in  \citet{CAB2010} (dashed curve) compared against our static comparison model A (solid curve). The parameters for the \crhydro\ model are shown in Table~\ref{table:static} and both of these models include MFA.
All spectra shown in Figure~\ref{fig:static_test} are static examples with no SNR evolution included and  $f_1(p)$ is measured at the subshock.
The superthermal portion of $f_1(p)$ below the high-energy turnover is extremely well matched in both shape and absolute normalization, as is the value of $\pmax$. 
We note that we determine the shape of the turnover at $\pmax$ differently than 
\citet{CAB2010}.
Following previous work 
\citep[e.g.,][]{EDB2004,LKE2008,ESPB2012}, we parameterize the turnover by multiplying $f(x,p)$ by an exponential factor, i.e.,
\begin{equation}
f(x,p) \rightarrow f(x,p) 
\exp{\left[-\frac{1}{\Bcut}\left(\frac{p}{\pmax}\right)^{\cutoff}\right]}
\ ,
\label{eq:rollover}
\end{equation}
%
%
%
where $\cutoff$ is a free parameter and $\Bcut=1$ or $\cutoff$ depending on the particular application. 
Results for different values of $\cutoff$ are shown and labeled in the top panel. In all of the cases shown in 
Figure~\ref{fig:static_test}, $\Bcut=\cutoff$.
The vertical solid lines indicate the $\pmax$ obtained from the \crhydro\  model. 

\begin{figure}
\centering
\includegraphics[width=8cm]{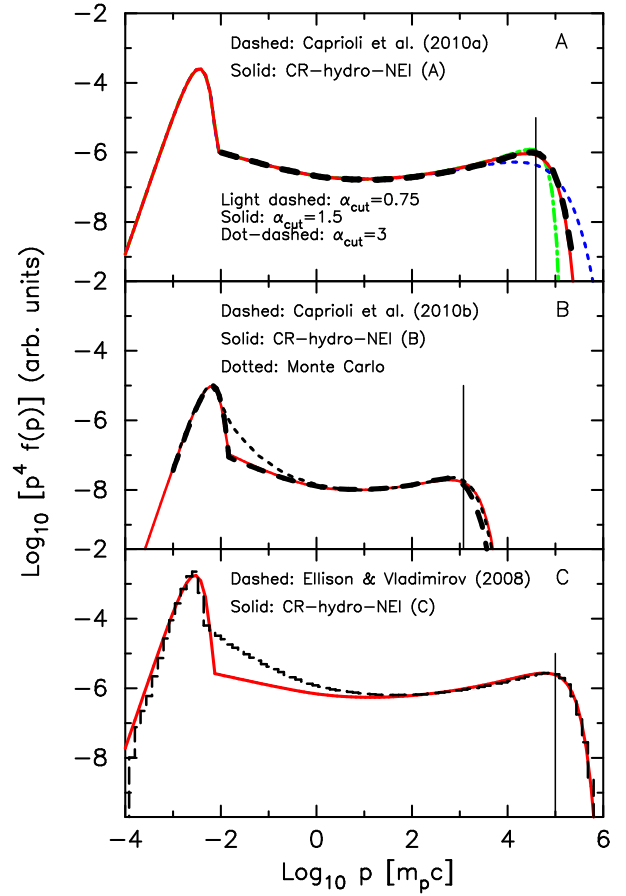}
\caption{Top panel - Comparison between our static Model A 
(light-weight colored curves) and the efficient acceleration example in 
\citet{CAB2010} (heavy black dashed curve). Both of these models include MFA. Curves of different colors correspond to different $\cutoff$ assumed: $\cutoff = 1.5$ (red solid), $\cutoff = 0.75$ (blue dashed) and $\cutoff = 3.0$ (green dash-dotted).
Middle panel - Comparison between our static Model B (solid red curve) and the benchmark model in \citet{CKVJ2010} (dashed curve - semi-analytic model). The dotted curve in the middle panel is the  Monte Carlo model from 
\citep{CKVJ2010}. These models do not include MFA.
Bottom panel - Comparison between \crhydro\ (model C) (solid red curve) and the \mc\ result from \citet{EV2008}. Both of these models include MFA.
The vertical solid lines indicate the $p_\mrm{max}$ obtained by our static models (see Section~\ref{sec:pmax}).
In the top panel, results with different cut-off indices ($\cutoff$ in equation~\ref{eq:rollover}) are shown as indicated. In the middle and bottom  panels, the \crhydro\ model assumes $\Inj=3.1$ and $3.5$ respectively, and $\cutoff=1.5$ for both. In all \crhydro\ results, $\Bcut=\cutoff$. 
}
\label{fig:static_test}
\end{figure}

While we could have adapted the procedure described in 
\citet{CAB2010} \citep[or other models such as][]{ZP2008} to determine the shape of the turnover, we feel the fundamental uncertainty in this shape is currently
best described by free parameters 
\citep[see][for a fuller discussion]{EB2011}. The CRs that escape upstream from the shock will have a highly anisotropic distribution and are certain to generate turbulence that will determine the nature of the FEB and the shape of the escaping population \citep[see, for example,][]{BOE2011}.  
Since the shapes of the trapped and escaping CR distributions,
at the highest accelerated energies, are critical for modeling both
X-ray synchrotron emission and GeV-TeV \gamray\ emission, we
feel it is important to have a parameterized model
that can be compared to observations until a more definitive theory of self-generated, long-wavelength turbulence  is developed.

A clear difference in the results  displayed in the top panel of
Figure~\ref{fig:static_test} is that we include the thermal population in 
$f_1(p)$. The thermal part of $f_1(p)$ is calculated assuming it is \MB\  with the shocked temperature and density that are \SCly\ determined by the DSA calculation.
A similar technique is used for the \SA\ calculation in 
\citet{CKVJ2010} and the full spectrum is shown in the middle panel for both the \citep{CKVJ2010} and \crhydro\ models.\footnote{We note that while the \SA\ solution determines the pressure accurately, the ``thermal" part of $f(x,p)$ does not have to be strictly \MB\ in shape immediately behind the shock. This will depend on the equilibration time scale.} 
The models in the top panel include MFA while those in the  middle panel do not.

The examples shown in the top two panels of 
Figure~\ref{fig:static_test}, and other tests  
we have performed but do not show, convince us that we have successfully included the static NL DSA calculation developed by P. Blasi and co-workers 
in the \crhydro\ code. Our implementation is not identical to that used in \citet{CAB2010} because we have chosen to treat some processes, such as CR escape and the calculation of the diffusion coefficient, in a different fashion. However, a suitable choice of  parameters can yield essentially identical results as 
Figure~~\ref{fig:static_test} demonstrates.

This in not the case for the \mc\ comparisons shown in the middle and bottom panels of Figure~\ref{fig:static_test}. The \mc\ simulation makes fundamentally different assumptions  from the \SA\ calculation, particularly  
 for how injection is treated and how the highest energy CRs escape from the shock. These differences show up in the spectra shown in Figure~\ref{fig:static_test}. In contrast to the \SA\ model, the \mc\ simulation does not make a diffusion approximation, i.e., it is never assumed that the particle distribution is nearly isotropic in some frame. Since thermal particles are never nearly isotropic in the shock frame, the injection of thermal particles cannot be directly modeled with a diffusion equation such as (\ref{eq:diffeq}). Escaping CRs will also have a highly anisotropic distribution making them hard to model within a diffusion approximation.

For the parameters  used for the ``benchmark case" in \citet{CKVJ2010}, the \mc\ model yields a high injection efficiency and $\Inj=3.1$ was chosen to match it for both of the \SA\ models in the middle panel. While this $\Inj$ yields an acceleration efficiency and total compression ratio that is  very similar for the \SA\ and \mc\ results, the spectrum in the transition region between thermal and 
\rel\ CRs is quite different. Of course, the \mc\ results depend on the assumptions of the model which do not include a description of the complex plasma processes that are certain to be important in the subshock layer.  The shapes of the spectra near $\pmax$ are very similar in the three cases shown in the middle panel where we have used $\cutoff=\Bcut=1.5$ for the \crhydro\ case 
\citep[see][for a thorough discussion of the shape of the turnover near $\pmax$]{CKVJ2010}.  

In the bottom panel of Figure~\ref{fig:static_test} we compare the \crhydro\ model to the \mc\ results presented in \citet{EV2008}. An injection parameter $\chi_\mathrm{inj} = 3.5$ is chosen to match the total compression ratio obtained by the \mc\ simulation 
($\Rtot \sim 9$). A similar difference in the transition region between thermal and superthermal  shows up in the two models as expected. Here, however, there are clear differences in the shapes of the spectra over a larger fraction of the momentum range. Particularly important is the difference in the location of the minimum in the $p^4 f(p)$ plot; the \MC\ minimum is at a momentum $\sim 5$  times as high as the \SA\ result. 

\begin{figure}
\centering
\includegraphics[width=8cm]{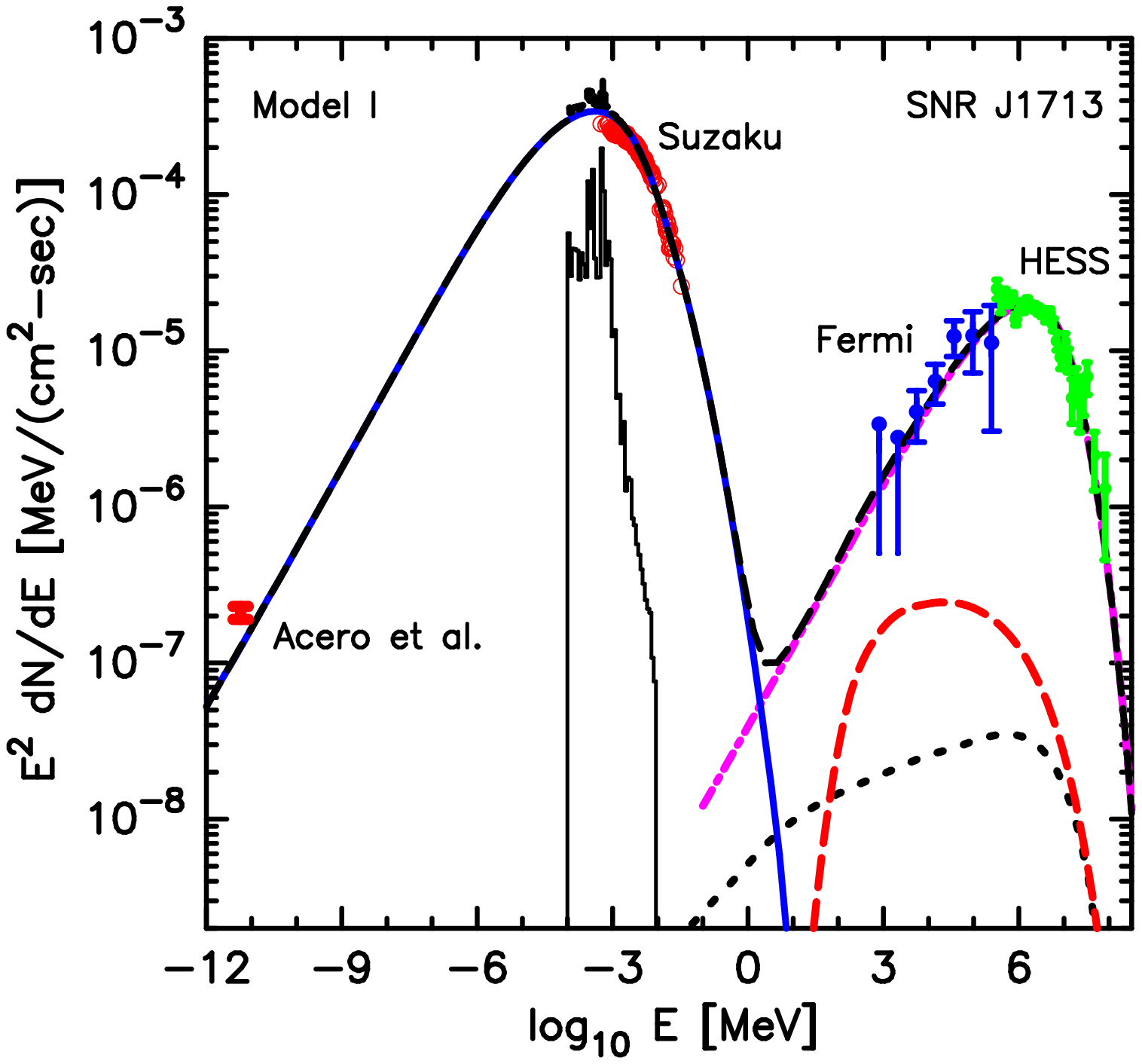} 
\caption{Broadband spectrum and the ``best-fit" model for 
SNR RX J1713.7-3946 listed in Table~\ref{table:param} as 
Model I.
The different emission
processes are: \syn\ (solid blue curve), IC (dot-dashed purple curve),
\pion\ from trapped CRs (dashed red curve), 
\brems\ (dotted black curve), 
and thermal X-rays (solid black curve). The
dashed black curve is the summed emission. The data is from
\citet{Acero2009} (radio), \citet{Tanaka2008} (\Suz\ X-rays),
\citet{AbdoJ1713_2011} (\Fermi), and
\citet{Aharonian_J1713_ERR2011} (\HESS). Note that the two lowest energy \Fermi\ points are upper limits. A column density of $n_H = 7.9\xx{21}$\,cm$^{-2}$ has been used for absorption and the model spectra have been multiplied by a factor  $\Norm=0.65$ to match the overall normalization of the observations. 
As noted in the text, the low-energy thermal X-ray emission is fully consistent with the \Suz\ data once the instrument response is considered.
%
}
\label{fig_rxj1713}
\end{figure}

\subsection{Pre-SN wind model of \SNRJ\ revisited}
\label{sec:J1713}
We now examine  how the modeling of 
\SNRJmm is modified using the generalized \crhydro\ code compared with a previous pre-SN wind model using the code described in \citet{ESPB2012} and references therein. 
A `best fit' result using our generalized code is shown in 
figures~\ref{fig_rxj1713} and \ref{fig:evo_J1713}   
(Model I in Table~\ref{table:param}) and it is clear, when 
Figure~\ref{fig_rxj1713} is compared to figure 2 in \citet{ESPB2012}, that our generalized model can yield a fit comparable in quality to that shown in \citet{ESPB2012}.
The most important conclusion from \citet{ESPB2012}, that \IC\ emission from electrons scattering off the CMB photons  dominates the GeV-TeV emission, remains robust.
As in \citet{ESPB2012}, the emission shown in Figure~\ref{fig_rxj1713} is the integrated emission from the interaction region between the CD and the FS and the shocked electron temperature is determined assuming the electrons are initially cold [i.e., $T_e = (m_e/m_p)T_p$] and then heated by Coulomb collisions, as discussed in 
Section~\ref{sec:TempEq}. 
In fact, in this case, the  dominance of IC over \pion\ at 
GeV-TeV energies doesn't depend on how electrons are heated although specific X-ray line ratios do. 
The fitting parameters used here for 
Figure~\ref{fig_rxj1713} are close enough to those used in 
\citet{EPSR2010} so the detailed discussion of the electron equilibration in that paper applies here. We refer the reader to text around Figures 1 and 2 in \citet{EPSR2010} for a full discussion.
The emission from the region between the \RS\ and the CD is ignored and the \pion\ emission from escaping CRs is not shown in 
Figure~\ref{fig_rxj1713} because we believe this will be insignificant for \SNRJmm\ for the reasons detailed in \citet{ESPB2012}.
\begin{figure}
\centering
\includegraphics[width=8cm]{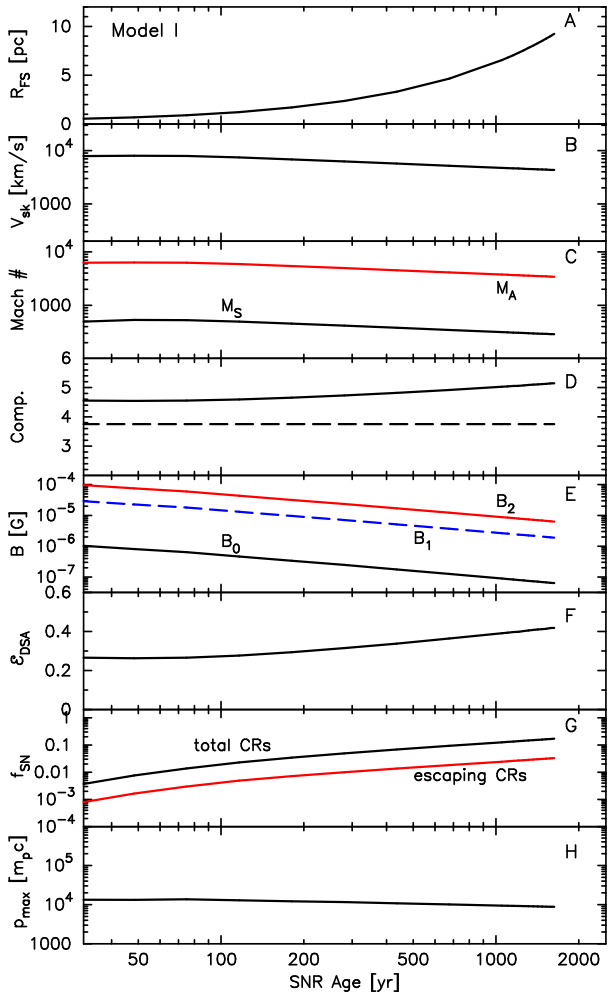}
\caption{Evolution of Model I for \SNRJmm. Panels A, B, and C show the FS radius, speed, and Mach numbers, respectively. In panel D the solid curve is the total compression ratio, $\Rtot$, and the dashed curve is $\Rsub$. The magnetic fields corresponding to far upstream, $B_0$, immediately upstream, $B_1$, and immediately downstream from the FS, $B_2$, are shown in panel E. 
Panel F shows the instantaneous DSA efficiency and panel G shows the fraction, $\fSN$,  of supernova explosion energy, $\EnSN$, put into all CRs (black curve) and only escaping CRs (red curve). The maximum particle momentum shown in panel H is essentially the same for protons and electrons since radiation losses are small for this example.
The transition from an age-limited to a size-limited $\pmax$ occurs at $t < 30$\,yr.}
\label{fig:evo_J1713}
\end{figure}

%
%

We note that, as in \citet{EPSR2010,ESPB2012}, we have chosen parameters for \SNRJmm\ that produce some thermal X-ray emission above the observations at the low-energy end of the \Suz\ energy range. While parameters could have been used that would produce no thermal emission above the \Suz\ observation, it must be kept in mind that the 
low-energy range is uncertain due to interstellar absorption and the \Suz\ instrument response. The model we show in Figure~\ref{fig_rxj1713} is fully consistent with the \Suz\ observations \citep[see][for a full discussion]{EPSR2010}. Any modification of the parameters to lower the low-energy thermal emission would strengthen the conclusion that IC dominates the GeV-TeV emission. This is also the case if any thermal X-ray emission from the RS is included.

It is important to note that, because of the number of parameters in the \crhydro\ model,  there is a range in each parameter (typically $\pm 20$\%) that allows a fit that is essentially 
equivalent to that shown in figure~\ref{fig_rxj1713} when other parameters are adjusted accordingly. To illustrate this, we list the parameters for Model~II in Table~\ref{table:param} that produce a
plot (not shown) almost indistinguishable from  Model I.  
Despite this freedom, there is no set of parameters that allows \pion\ to dominate IC at GeV-TeV energies as long as the constraint from the thermal X-rays emission lines is considered. The low thermal X-ray luminosity forces the ambient density to be low and this results in  IC dominance at \gamray\ energies.
While the fit to \SNRJmm\ shown here is similar to that shown in \citet{ESPB2012}, there are differences in the fitting parameters stemming from the way MFA, the CR diffusion coefficient, $\pmax$, the shock precursor, and other aspects of NL DSA are calculated.

In the model shown in figure 2 of \citet[][]{ESPB2012}, the acceleration efficiency was set at $\EffDSA=25$\% and the injection parameter, $\Inj$, was determined as the SNR evolved to produce this percentage.\footnote{The acceleration efficiency, $\EffDSA$, is the percentage of far upstream energy flux crossing the shock at any given time (in the shock rest frame) that ends up in superthermal particles.} 
Here, we fix $\Inj=3.75$ and the acceleration efficiency is determined directly from this.\footnote{This change is implemented mainly to speed computation. In our current \crhydro\ code, runs with fixed $\EffDSA$ take significantly longer to run than those with fixed $\Inj$.} 
In both cases, either $\EffDSA$ or $\Inj$  was chosen to obtain a good fit to the observations. With $\Inj=3.75$, the instantaneous acceleration efficiency varies smoothly
between $\EffDSA\simeq 25$\% early in the evolution 
(i.e., $\tSNR\simeq 200$\,yr), and $\EffDSA \simeq 35$\% at the end of the run ($\tSNR=1630$\,yr). 
At $\tSNR=1630$\,yr, the  fraction of SN explosion energy, $\EnSN$, that has been placed into CRs, both trapped and escaping, is 
$\sim 16$\%. In the \citet{ESPB2012} fit, $\sim 15$\% of $\EnSN$ was put into CRs by 1630\,yr. We emphasize that although IC emission from electrons dominates the broadband emission, the SNR forward shock is accelerating CR ions with the efficiencies just quoted and the CR electrons carry a much smaller fraction of the energy.

In \citet{ESPB2012} an `ad hoc' amplification factor, $\Bamp=8.5$,  was used where the compressed  magnetic field at the shock was simply multiplied by a factor chosen to obtain a fit to the observations. Here, the MFA is calculated in a more \SC\ manner as described in Section~\ref{sec:model}. In both cases, however, parameters are chosen to produce a shocked magnetic field of $\sim 10$\,\muG, as required to fit the observations. 
The shocked magnetic field depends on a number of parameters including 
$\SigWind=B(r)^2/[4 \pi \rho(r) \Vwind^2]$ 
\citep[][]{CL94} and the amount of damping parameterized with $\fDamp$. For Model~I, $\SigWind=0.004$ and $\fDamp=0.1$, while for Model~II, 
$\SigWind=0.008$ and $\fDamp=0.3$. As shown in panel E of 
Figure~\ref{fig:evo_J1713}, the pre-SN wind magnetic field, $B_0$ is increased by $\sim 100$ fold at $\tSNR=1630$\,yr.
As mentioned in \citet{ESPB2012}, the low magnetic field values we find are quite different from the high-field estimates 
($\sim 1$\,mG) obtained for \SNRJmm\ from  rapid time variations and sharp X-ray edges in filaments \citep[e.g.,][]{Uchiyama_J1713_2007}.
However,
other estimates yield lower values 
\citep[see references in][]{EV2008}, and \citet{BykovDots2008} 
provide  an alternative explanation for
rapid time variations that does not require such large fields.
Furthermore, our estimate is for the integrated magnetic field strength in the CD-FS interaction region and local values in small regions could be much higher than our average value.

An important parameter for DSA in general, and a critical one for determining the relative importance of IC vs. pion-decay at GeV-TeV energies, is the electron to proton ratio at 
relativistic energies, $\Kep$. This ratio has not yet been reliably determined from theory or plasma simulations and a wide range of values, i.e., $0.02\, \gsim \Kep \gsim 10^{-4}$, have been used to fit SNR J1713 \citep[see, for example,][and references 
therein]{KW2008,MAB2009,ZirA2010,ESPB2012}. Values of $\Kep \ll 0.01$ allow pion-decay to dominate the gamma-ray emission but are not consistent with  the constraint imposed by the weak thermal X-ray luminosity. Here, as in our previous work, we find that $\Kep \sim 0.01$ provides an excellent fit to the broad-band emission and that values $\Kep \ll 0.01$ can be excluded.

The \SA\ DSA model used in \citet{ESPB2012} assumed the CR diffusion coefficient was a rapidly increasing function of momentum without explicitly  specifying the momentum or spatial dependence. Our generalized model used to produce the fit shown in 
Figure~\ref{fig_rxj1713}, uses the explicit form given in 
equation~(\ref{eq:diff}) with $\alpha=1$ and a spatial dependence determined by $B(x)$.

The determination of $\pmax$ is also more consistently done here than in \citet{ESPB2012}, as described in Section~\ref{sec:pmax}. However, when the size of the shock determines $\pmax$, as is the case for Figure~\ref{fig_rxj1713}, there still must be a free parameter, $\fFEB$, relating the finite shock radius to $\pmax$. Here, $\fFEB=0.2$ while $\fFEB=0.1$ was used for the fit in figure 2 of \citet{ESPB2012}.
There are a number of other minor differences as seen by comparing Model A in Tables 1-4 in \citet{ESPB2012} to Model I in 
Tables~\ref{table:param} and \ref{table:ModelOut} here.

In Figure~\ref{fig:evo_J1713} we show the evolution of some parameters over the lifetime of \SNRJmm. At $\tSNR=1630$\,yr, the magnetic field in the pre-SN wind has dropped to $B_0 \sim 10^{-7}$\,G and the field just behind the FS, $B_2$, is $\sim 50$ times larger. 
The instantaneous acceleration efficiency, $\EffDSA$ (panel F), and the FS compression ratio $\Rtot$ (panel D) increase  
noticeably but over the brief the age of this remnant, $\pmax$ is essentially   unchanged at $\sim 10^4$\,$m_p c$.

\section{Conclusions}
We have performed an extensive generalization of a code (\crhydro) that models the evolution of a spherically symmetric SNR undergoing efficient, 
\NL\ DSA.  The \SA\ model for DSA that we use is based  on previous work by P. Blasi and co-workers \citep[e.g.,][]{CAB2010} and here we present  a detailed discussion of the DSA mechanism and how it is incorporated in the  hydrodynamic model of an evolving SNR. 

Relativistic particle populations are the signature of violent, nonthermal activity in astrophysical sources. These sources are ubiquitous and many of these populations are undoubtedly produced by collisionless shocks. 
Supernova remnants offer a promising platform for 
studying the shock acceleration process because they are well studied in a broad wavelength band, there are a number of different remnants in different environments,  it is certain that 
SNR shocks produce \rel\ electrons, and because it is likely that SNRs are the main sources of CR ions up to at least $\sim 10^{15}$\,eV.

As is clear from our discussion in Section~\ref{sec:model}, efficient DSA is intrinsically complicated  and there are a number of parameters required to properly model the NL  couplings between the SNR hydrodynamics, MFA, CR acceleration, and the \NEI\ production of thermal X-ray emission.
Given these complexities, our aim is to develop as complete a model as possible that can be used to interpret the broadband continuum and line emission from remnants in a reasonably \SC\ fashion. 

The most important generalizations included in \crhydro\ are  explicit calculations of the shock precursor structure and MFA. These effects are coupled to a momentum and position dependent CR diffusion coefficient, the calculation of the maximum particle momentum, $\pmax$, and a finite \alf\ speed for particle scatterers. Other additions include the ability to accelerate a superthermal seed population of CRs in addition to thermal particles and wave damping due to partially-ionized material.
We have performed a number of tests to ensure that we 
accurately reproduce the \SA\ results of \citet{CAB2010,CKVJ2010} where applicable, and we have noted where our model of NL DSA differs from that work. We have also compared our results to \mc\ simulations and discussed important differences of this alternate DSA model.

There are two main  features that distinguish \crhydro\ from other SNR models. The first is that we couple the remnant dynamics to the CR acceleration and the second is that we \SCly\ include a NEI calculation of thermal X-ray emission with the dynamics and CR acceleration. It turns out that the thermal X-ray emission is the key factor in determining the dominant GeV-TeV emission mechanism in \SNRJmm. The absence of thermal X-ray emission lines in this remnant forces the ambient density to be low enough that \IC\ emission from \rel\ electrons dominates  \pion\ from \rel\ ions. Without a consistent description of the thermal X-ray emission, parameters can be found for \SNRJmm\ with either mechanism dominating the GeV-TeV 
emission \citep*[e.g.,][]{MAB2009}. It is important to note, however, that depending on the ambient conditions, it is possible that the \gamray\ emission in other SNRs is dominated by \pion, as claimed for Tycho's SNR \citep[i.e.,][]{MC2011}, is some near-equal mix of IC and \pion, as suggested for CTB 109 (G109.1-1.0) \citep[][]{CSEP2012}, or is even \brems\ \citep[e.g.,][]{bceu00}.

In the interest of conciseness, we have not shown examples here with a superthermal seed population of CRs, ones showing the effects of ion-neutral damping in partially-ionized material, or examples where Coulomb losses are important for low energy CRs. These processes are expected to be important for SNRs interaction with dense material and in future work we will use \crhydro\ to  model the interaction of a remnant FS with a dense molecular cloud. 

\acknowledgments 
The authors acknowledge important discussions with A. Vladimirov and D. Patnaude concerning this work. D.C.E. acknowledges support from NASA grants
ATP02-0042-0006, NNH04Zss001N-LTSA, 06-ATP06-21, and NNX11AE03G. S.N. acknowledges support from Ministry of Education, Culture, Sports, Science and Technology (No. 23105709), Japan Society for the Promotion of Science (No. 19104006 and No. 23340069), and the Global COE Program, The Next Generation of Physics, Spun from Universality 
and Emergence, from MEXT of Japan.  

\begin{center} 
\begin{deluxetable*}{lcccccl} 
\tablecolumns{6}
\tablewidth{17.5cm}
\tablecaption{Model Parameters - Input \tnac}
\tablehead{\colhead{} & \colhead{Model \Rmnum{1} \tna} & \colhead{Model \Rmnum{2} \tna} & \colhead{Model \Rmnum{3}} & \colhead{Model \Rmnum{4}}  &\colhead{Model \Rmnum{5}} &\colhead{Remarks}}
\startdata
%
$\tSNR$ [yr]	& 1630	& 1630	& 1000	& 1000	& 1000	& SNR age\\
$\Inj$	& 3.75	& $3.75$ 	& $4.70$ 	& $3.60$ 	& $3.60$ 	& Injection parameter\\
$\fDamp$			& 0.1		& $0.3$ 	& $0.1$ 	& $0.1$ 	& $0.99$ 	& Wave-damping factor\\
$\alpha$			& 1.0		& $1.0$ 	& $1.0$ 	& $1.0$ 	&  $1.0$ 	&PL index of $D(x,p)$ \tnb\\
$\Kep$		& 0.01	& $0.01$ 	& $0.01$ 	& $0.01$ 	& $0.01$ & (e$^-$/p) ratio at  $\pmax$ \tnc\\
$\cutoff$	& 0.5		& 0.5		& 1.0		& 1.0		& 1.0 & Exp. cut-off index \tnd\\
$\Bcut$	& 1		& 1		& 1		& 1		& 1 & Exp. cut-off factor \tnd\\
$\falf$	& 0.25		& 0.25		& 0.0		& 0.0		& 0.0 & $x$-dependance of $\Valf$ \tne\\
$\fFEB$	& 0.20		& 0.20		& 0.20		& 0.20		& 0.20 & $L_\mrm{feb} / R_\mrm{FS}$\\
$\dMdt$ [\SunMyr]	& $7.5\xx{-6}$		& $7.5\xx{-6}$	&  $-$		& $-$		& $-$ & Pre-SN Mass loss \\
$\Vwind$ [\kmps]	& 20 	& 20 &  $-$		& $-$		& $-$ & Pre-SN wind speed \\
$\SigWind$	&  0.004		& 0.008	&  $-$		& $-$		& $-$ & Wind magnetization \\
$\Nphot$	& 1		& 1 		& $-$		& $-$ 		& $-$  & No. of seed photon fields \tnf\\
$\Norm$	& 0.65		& 0.70 		& $-$ 		& $-$ 		& $-$ 	 & Flux  normalization  \tng
%
\enddata
\tablenotetext{$\ast$}{For all models listed in this Table, MFA is included and the fraction of helium by number is taken to be $\fHe=0.0977$.}
\tablenotetext{a}{Best-fit models for \SNRJmm. These models all have $\EnSN=10^{51}$\,erg, $\Mej=3\,\Msun$, 
$\tSNR=1630$\,yr, $\dSNR=1$\,kpc and $\Twind=10^4$\,K.}
\tablenotetext{b}{See e.g., equation~(\ref{eq:diff}). Here, $\alpha = 1$ refers to Bohm diffusion, for which we set $D(x,p) = pcv(p)/[3e \delta B(x)]$. 
}
%
\tablenotetext{c}{If radiation losses are important for electrons, $\Kep$ is measured  at \rel\ energies but below the cutoff.}
\tablenotetext{d}{See equation~(\ref{eq:rollover}).}
\tablenotetext{e}{See equation~(\ref{eq:vAlf}).}
\tablenotetext{f}{$\Nphot=1$ means only the CMB photons are considered for IC emission.}
\tablenotetext{g}{Values other than 1 can reflect uncertainties in the distance and/or the fraction of FS surface that accelerates CRs.}
\label{table:param}  
\end{deluxetable*}
\end{center}
%
\begin{center}
\begin{deluxetable*}{lcccccl} 
\tablecolumns{7}
\tablewidth{18.0cm}
\tablecaption{Model Parameters - Output\tnaa}
\tablehead{\colhead{} & \colhead{Model \Rmnum{1}} & \colhead{Model \Rmnum{2}} & \colhead{Model \Rmnum{3}} & 
\colhead{Model \Rmnum{4}} & \colhead{Model \Rmnum{5}} & \colhead{Remarks}}
\startdata
$\Msonic$		& 291 	&  291	& 251 	& 219 	& 220	& Sonic Mach number\\
$\Malf$		& 2190	& 1550 	& 216 	& 188 	& 189	& \alf\ Mach number\\
$\VelFS$ [km/s]  &$4.38\xx{3}$		&$4.39\xx{3}$		& $3.78\xx{3}$ 	& $3.30\xx{3}$ 	& $3.32\xx{3}$ & FS speed\\
$\RFS$ [pc] 		& $9.31$ 	&$9.30$  	& $6.91$ 	& $6.38$ 	& $6.45$	& FS radius\\
$\EffDSA [\%] $		& 39.7	& 38.8 	& $3.46$ 	& $83.8$ 	& $79.6$	& Total DSA efficiency\\
$\EffEsc [\%]$		& 7.60	& 7.43 	& $3.43$ 	& $32.4$ 	& $27.9$	& Efficiency for escaped CRs\\
$\Rtot$		& 5.05 	& 5.01 		& $4.00$ 	& $9.20$ 	& $8.39$	& Total comp ratio\\
$\Rsub$		& 4.00 	& 3.99 		& $4.00$ 	& $3.63$ 	& $3.68$	& Subshock comp ratio\\
$B_2$ [\muG]	& 6.71 	& 6.91 		& $10.1$ 	& $154$ 	& $17.5$	& B-field right behind 
FS\\
$\pmax$ (p) [$m_p c$]	& $9.35\xx{3}$ 	&$9.59\xx{3}$ 	& $1.22\xx{3}$ 	& $1.19\xx{5}$ 	& $1.19\xx{4}$ & Max proton momentum\\
$\pmax$ (e) [$m_p c$] 	& $9.35\xx{3}$	&$9.59\xx{3}$ & $1.22\xx{3}$	& $9.21\xx{3}$ 	& $1.19\xx{4}$ & Max electron momentum\\
$\crPup(x=0)$	& 0.21	& 0.20 	& $1.17\xx{-4}$ 	& $0.64$ 	& $0.55$ 	& CR pressure at subshock\\
$(P_w/\Pgas)_1$	& 0.78	& $0.36$ 		& $2.60\xx{-2}$ 	& $3.86$ 	& $4.81\xx{-3}$ & Wave/gas pres. at subshock\\
$\crPgas$			& 0		& 0 			& 0 		& 0 		& 0 	& Pressure in pre-existing CRs
%
%
\enddata
\tablenotetext{$\dagger$}{Output values at the end of the simulation ($t = \tSNR$).}
\label{table:ModelOut}  
\end{deluxetable*}
\end{center}
%
\begin{center}
\begin{deluxetable*}{lccccccccc}  
\tablecolumns{10}
\tablewidth{16.0cm}
\tablecaption{Static Model Parameters\tnab}
\tablehead{\colhead{} & \colhead{$R_\mathrm{FS}$ [pc]} & \colhead{$M_S$} & \colhead{$M_A$} &  \colhead{$v_\mathrm{sk}$ [km/s]} & \colhead{$B_0$ [$\mu$G]} & \colhead{$n_0$ [cm$^{-3}$]} & \colhead{$T_0$ [$10^5$~K]}  & \colhead{$\chi_\mathrm{inj}$} & \colhead{MFA}}
\startdata
Model A\tna & $1.5$ & $135$  & $150$ & $5000$ & $5.0$ & $0.1$ & $1.0$ & $3.30$ & yes\\
Model B\tnb & $0.1$ & $30$ & $42$ & $5000$ & $3.0$ & $0.003$ & $20.0$ & $3.10$ & no\\
Model C\tnc & $1.0$ & $30$ & $140$ & $3000$ & $10.0$ & $1.0$ & $7.3$  & $3.50$ & yes 
\enddata
\tablenotetext{$\ddagger$}{For all models in this Table, $\cutoff =\Bcut=1.5$, $\fHe=0.0$, and $\fFEB=0.1$. For  Model A, the additional cases  $\cutoff (= \Bcut) = 0.75$ and $3.0$ are plotted.} 
\tablenotetext{a}{Comparison with \citet{CAB2010}.}
\tablenotetext{b}{Comparison with \citet{CKVJ2010}.}
\tablenotetext{c}{Comparison with \citet{EV2008}.}
\label{table:static}  
\end{deluxetable*}
\end{center}

\bibliographystyle{aa} 
\bibliography{reference_rev}

\end{document}